\input phyzzx
\hoffset=0.375in
\overfullrule=0pt

\def\kmsmpc{{\rm km}\,{\rm s}^{-1}\,{\rm Mpc}^{-1}}
\twelvepoint
\font\bigfont=cmr17
\centerline{\bigfont Cepheid Luminosity Versus Galaxy Rotation Speed: 
$L\propto v^{0.7}$}
\bigskip
\centerline{{\bf Andrew Gould}\footnote{1}{Alfred P.\ Sloan Foundation Fellow}}
\smallskip
\centerline{and}
\smallskip
\centerline{\bf Piotr Popowski}
\smallskip
\centerline{Dept of Astronomy, Ohio State University, Columbus, OH 43210}
\smallskip
\centerline{e-mail gould@payne.mps.ohio-state.edu, 
popowski@payne.mps.ohio-state.edu}
\bigskip
\centerline{\bf Abstract}
\singlespace 
The distance modulus of a galaxy determined from Cepheids differs from its
distance modulus determined from Tully-Fisher by an amount that is proportional
to the galaxy's line width, $W_R^i$: 
$\Delta\mu \propto (-1.73\pm 0.46)\log W_R^i$.
While a miscalibration of the slope of Tully-Fisher could in principle 
produce this effect, we argue that such a miscalibration is very unlikely.  
The other possible explanation is that the inferred Cepheid luminosity is 
correlated with the rotation speed $v$ 
(and hence $W_R^i$) of its parent galaxy: $L\propto v^{0.7}$.  
(This proportionality is superposed on the well-established relation between 
period and luminosity.)\ \ Such a dependence would be
expected if Cepheid luminosity
is correlated with metallicity, since galaxies with deeper potential wells tend
to retain more of 
their metals.  It would induce a discrepancy of 0.83
mag between $H_0$ determinations that are calibrated using M31 
($\log W_R^i=2.71$) and the $H_0$ determination from SNIa, which is calibrated
in galaxies with mean line width $\VEV{\log W_R^i}=2.23$.  A discrepancy of
0.83 mag corresponds to the difference between $80\,\kmsmpc$ and $55\,\kmsmpc$,
very similar to the actual discrepancy reported in the current literature.
\bigskip
Subject Headings: Cepheids -- galaxies: distances -- stars: early type
\smallskip
\centerline{submitted to {\it The Astrophysical Journal Letters}: 
February 24, 1997}
\centerline{Preprint: OSU-TA-4/97}
\endpage
\chapter{Introduction}

	Cepheids are the most luminous primary distance indicators.  As such,
they provide the only method to make direct measurements of the distance
to galaxies outside the Local Group.  Even so, prior to this decade 
Cepheid measurements were limited to galaxies within a few Mpc.  With the
advent of the {\it Hubble Space Telescope (HST)} Key Project (Freedman 1997
and references therein), however, Cepheid distance determinations are being
obtained for galaxies as far as the Virgo and Fornax clusters.  The plan
of the Key Project is to calibrate and so cross-check many secondary distance
indicators.

	A major potential problem for this program is that the inferred 
Cepheid 
luminosities (and so the inferred distances) may be a function of metallicity.
If so, one could in principle correct the Cepheid distances by determining the
correlation of inferred distance with metallicity and then measuring the
metallicity of the individual galaxies in which the Cepheids lie.  In practice,
it is difficult to obtain reliable metallicities for distant galaxies.
Freedman \& Madore (1990) argued that there is no significant correlation 
between metallicity and inferred distance based on their analysis of
Cepheids found in three
fields of varying metallicity in M31.  However, Gould (1994) re-analyzed the 
same data using a more rigorous statistical procedure and found a significant
dependence over the metallicity range $-0.5<\rm [Fe/H]<0.2$.  
Depending on assumptions about systematics, he measured a slope of either
$${\Delta \mu\over \Delta {\rm [Fe/H]}}= 0.88\pm 0.16,\quad{\rm or}\quad
{\Delta \mu\over \Delta {\rm [Fe/H]}}= 0.56\pm 0.20
,\eqn\gouldeq$$
where $\Delta \mu/ \Delta\rm [Fe/H]$ is the correction to the apparent distance
modulus per dex of metallicity.  The theoretical work of Stothers
(1988) and Stift (1995) predict effects of the same order.

	Gould (1994) argued that if this relation continued to lower 
metallicities, it could well explain the discrepancy between different
determinations of the Hubble Constant $(H_0)$.  For example, the calibration
of type Ia supernova is based on Cepheid measurements of metal-poor galaxies
(whose distances would be relatively overestimated) while the methods of 
surface brightness fluctuations and planetary nebulae are calibrated in M31
(whose distance would be relatively underestimated).  However, the one test
that was then available, comparison of the Cepheid and RR Lyrae distances to
IC1613, did not appear to confirm the trend.

	Subsequently, Beaulieu et al.\ (1997) found a somewhat smaller
dependence ($\Delta \mu/ \Delta\rm [Fe/H]=0.44_{-0.2}^{+0.1}$)
by comparing Cepheids in the Large and 
Small Magellanic Clouds.  They also suggested that much of the 
conflict between different determinations of $H_0$ could be resolved by 
taking this effect into account.

	Here we show that the ratio of the distance of a galaxy as determined
from Cepheids to its distance as determined from Tully-Fisher is a strong
function of the galaxy's rotation speed $v$ (as inferred from its line width
$W_R^i$):
$${D_{\rm Ceph}\over D_{\rm TF}}\propto v^{-0.35}.\eqn\cephtf$$
We argue that this dependence is not likely to be due to a miscalibration
of the slope of the Tully-Fisher relation. We therefore conclude that 
Cepheid distances as presently calibrated deviate systematically from the
true distances by an amount that depends on the rotation speed of the galaxy.
Such a correlation is plausible because galaxies with deeper potential wells
retain more of their metals and, according to equation \gouldeq, Cepheids of
higher metallicity are brighter.  We therefore interpret equation \cephtf\ as
implying
$$L_{\rm Ceph}\propto v^{0.7}.\eqn\cephlum$$
This proportionality is valid at fixed period and is superposed on the 
well-established relation between period and luminosity.

	If confirmed by continued investigations, the correlation expressed
in equation \cephlum\ could prove extremely useful.  It would mean that
rotation speed, which is usually quite easy to measure, could be used as a 
proxy for metallicity in determining the correction to Cepheid distances.  

\chapter{Data}

	Shanks (1997) finds the zero-point of Tully-Fisher is $0.46\pm 0.14$
mag  brighter when calibrated from 11 galaxies with {\it HST} Cepheid 
distances than it is based on the traditional ground-based calibration.
Here we repeat Shanks' analysis but with two changes.  First, Shanks 
effectively fits the data to a model with two parameters: the zero-points of 
the {\it HST}-calibrated and ground-calibrated Tully-Fisher relations.  To
these we add a third parameter, the line width of the galaxy corrected
for inclination, $W^i_R$.  Second, for our ground-based sample we restrict
attention to the four galaxies with good $BVRI$ Cepheid distances, whereas
Shanks considers six galaxies.  We do so in order to make this sample 
(like the {\it HST} sample) as homogeneous as possible and in particular to
minimize errors that might be introduced by poorly determined extinction.

\FIG\one{
Difference ($\Delta \mu$) between distance moduli as determined from Cepheids
and Tully-Fisher as a function of galaxy line width.  Open circles are for
{\it HST} Key Project galaxies and solid triangles are for local Tully-Fisher
calibrators with $BVRI$ data.  Solid lines are a 3-parameter
fit: one zero-point for the {\it HST} Cepheids, one for the ground-based 
Cepheids, and a single slope for both.  The dashed line is a 2-parameter fit
with one zero-point for all galaxies.
}

	Figure \one\ shows the distance-modulus difference 
($\Delta\mu=5\log D_{\rm Ceph}/D_{\rm TF}$) versus 
$\log W_R^i$ for 15 galaxies.  The
open circles are taken from Table 1 of Shanks (1997).  The error bars are 
determined by adding the Cepheid and Tully-Fisher errors in quadrature.
The solid triangles 
represent the distance modulus difference of ground-based $BVRI$ 
Cepheid distances with Tully-Fisher distances.  In order of increasing
line width, the Cepheid distances for NGC300, M33, M81, and M31, are taken
from
Freedman et al.\ (1992),
Freedman, Wilson, \& Madore (1991),
Madore, Freedman, \& Lee (1993),
and
Madore \& Freedman (1991).
The Tully-Fisher distances to these four galaxies
are derived using the calibration,
line widths, $B$ magnitudes, and internal and external extinction from
Pierce \& Tully (1992).  The Tully-Fisher errors are taken to be 0.30 mag and 
these are again
added in quadrature to the Cepheid errors as reported by the observers.

	We fit the data to the form
$$\Delta\mu = \alpha + \beta (\log W^i_R - 2.5) + \delta_{HST},\eqn\fitform$$
where $\delta_{HST}$ is defined to be zero for the four local calibrator 
galaxies.  We find that
$$\alpha = 0.10\pm 0.17,\quad \beta=-1.73\pm 0.46,\quad 
\delta_{HST}=0.35\pm 0.20,
\eqn\bestfits$$
and $\chi^2=6.37$ for 12 degrees of freedom.  Note that the slope $\beta$
is detected at the $4\,\sigma$ level so that from a statistical point of view
it is certainly justified to introduce this third parameter.

	The offset $\delta_{HST}$ is detected only at the $2\,\sigma$ level,
so the existence of an offset may or may not be a real effect.  If this 
parameter is removed from equation \fitform, we find a best fit of
$$\alpha = 0.34\pm 0.09,\qquad \beta=-1.73\pm 0.46, 
\eqn\bestfitsp$$
and $\chi^2=9.48$ for 13 degrees of freedom.  That is, the slope is 
unaffected by assumptions about the existence of an offset.  To the eye it
may appear that the point at far upper left (IC4182) plays an unreasonably
large role in the fit.  However, even if this is removed, the best fit 
parameters change by well under $1\,\sigma$.  In particular the slope is
$\beta=-1.68\pm 0.59$.

\chapter{Discussion} 

	Barring an extreme statistical fluctuation, the slope
$\beta=-1.73$ found in the previous section must be due to systematic
errors in either Tully-Fisher distances, Cepheid distances or both.  
If, for example, the slope of the Tully-Fisher relation had been improperly
calibrated, and it was actually 1.73 higher ($-5.75$ rather than $-7.48$ in 
$B$), then the entire effect seen in Figure \one\ would be explained.

	Is such a large error in the slope of Tully-Fisher possible?  This
slope is determined from a group of galaxies in the Ursa Major cluster
by plotting their apparent magnitudes against their line widths.
A quick glance at Figure 3 from Pierce \& Tully (1992) shows that the fit is
extremely good.  The Ursa Major galaxies are assumed all to be at the same
distance, but even if they were not, this would only increase the scatter
on the diagram and would 
not change the slope.  The one remaining possible problem
with Tully-Fisher would be if its slope in Ursa Major differed substantially 
from that of the galaxies plotted in Figure \one.  This solution
appears highly implausible to us.

	The alternative explanation is that Cepheid luminosities (and therefore
inferred Cepheid distances) depend on the rotation speed of the galaxy in
which they happen to lie.  As discussed in \S\ 1, a dependence with this sign
is expected if the inferred Cepheid luminosity (taking account of the
apparent extinction as inferred from observed colors) rises with increasing
metallicity.  Although metallicities are not available for most of the sample,
we estimate very roughly that over the range of line widths probed
($\Delta \log W_R^i\sim 0.8$) the metallicity varies by 1.5 dex.  We make
this estimate by assuming that for the lowest line widths, [Fe/H]$\sim -1.3$
(Beaulieu et al.\ 1997 and references therein), and for the highest line
widths, [Fe/H]$\sim 0.2$ (Freedman \& Madore 1990 and references therein).
This leads
to an estimate ${\Delta \mu/ \Delta {\rm [Fe/H]}}\sim 1.15\pm 0.31$, in 
qualitative
agreement with the range of values found by Gould (1994) and reproduced in
equation \gouldeq.  We conclude that there is a strong case that the slope
seen in Figure \one\ is due primarily to a correlation between Cepheid 
distances and galaxy line widths which is rooted in the dependence of 
inferred Cepheid luminosity on metallicity.

	As discussed by Gould (1994) and Beaulieu et al.\ (1997), if
Cepheid distances depend on metallicity, then some of the divergent estimates
of the Hubble constant can be reconciled.  For example, surface brightness
fluctuations (Tonry et al.\ 1997) and planetary nebula (Jacoby 1996) which are
fundamentally calibrated in M31 ($\log W_R^i=2.71$) yield
$H_0\sim 80\,\kmsmpc$, while supernova type Ia (Sandage et al.\ 1996) which are
calibrated
in 7 galaxies with mean $\VEV{\log W_R^i}=2.23$ yields
$H_0\sim 55\,\kmsmpc$.  That is, they differ by 0.81 mag.  The discrepancy
that we would expect based on equation \bestfits\ is 
$1.73\times(2.71 -2.23)=0.83\,$mag.

	Finally, we note that while the evidence that we find for an offset
between the local and {\it HST} calibrations of Tully-Fisher is not as strong
as that reported by Shanks (1997), we do think that this question warrants 
further investigation.  We can think of two effects that might give rise to
such an offset.  First, it is possible that the {\it HST} Cepheids come
preferentially from more outlying (and so metal-poorer) regions of galaxies
than do the ground-based Cepheids.  This would cause them to be systematically
fainter for galaxies of the same line width.  Second, environment may affect
either the Tully-Fisher or the Cepheid determinations.  The {\it HST} Cepheids
come preferentially from clusters and of the two galaxies that appear to lie
closer to the ground-based triangles in Figure \one, one is a field galaxy and
the other is from the relatively quiescent Leo Group.  

{\bf Acknowledgements}:
This work was supported in part by grant AST 94-20746 from the NSF and in
part by grant NAG5-3111 from NASA.

\bigskip

\Ref\beau{Beaulieu, J.\ P., et al.\ 1997, A\&A, submitted (= astro-ph/9612215)}
\Ref\Fre{Freedman, W.\ L\ 1997, in Proceedings of ``Critical Dialogs in
Cosmology'', preprint (=astro-ph/9612024)}
\Ref\FM{Freedman, W.\ L., \& Madore, B.\ F.\ 1990, ApJ, 365, 186}
\Ref\nth{Freedman, W.\ L., Madore, B.\ F., Hawley, S.\ L., Horowitz, I.\ K.,
Mould, J., Navarrete, M., \& Sallmen, S.\ 1992 ApJ, 396, 80}
\Ref\mtt{Freedman, W.\ L., Wilson, C.\ D., \& Madore, B.\ F.\ 1991, ApJ, 372, 
455}
\Ref\gou{Gould, A.\ 1994, ApJ, 426, 542}
\Ref\jac{Jacoby, G.\ H., 1997, in {\rm The Extragalactic
  Distance Scale}, eds., M.\ Livio \& M.\ Donahue, in press}
\Ref\mto{Madore, B.\ F., \& Freedman, W.\ L.\ 1991, PASP, 103, 933}
\Ref\meo{Madore, B.\ F., Freedman, W.\ L., \& Lee, M.\ G.\ 1993, AJ, 106, 2243}
\Ref\pie{Pierce, M.\ J., \& Tully, R.\ B.\ 1992,  ApJ, 387, 47}
\Ref\san{Sandage, A., Saha, A., Tammann, G.\ A., Labhardt, L., Panagia, 
N., \& Macchetto, F.\ D.\ 1996, ApJ, 460, L15}
\Ref\sha{Shanks, T.\ 1997, preprint (=astro-ph/9702148)}
\Ref\Sto{Stift, M.\ J.\ 1995, A\&A, 301, 776}
\Ref\Sto{Stothers, R.\ B.\ 1988, ApJ, 329, 712}
\Ref\ton{Tonry, J.\ L., Blakeslee, J.\ P., Ajhar, E.\ A., \& Dressler, A.\
ApJ, 475, 399}
\refout
\figout
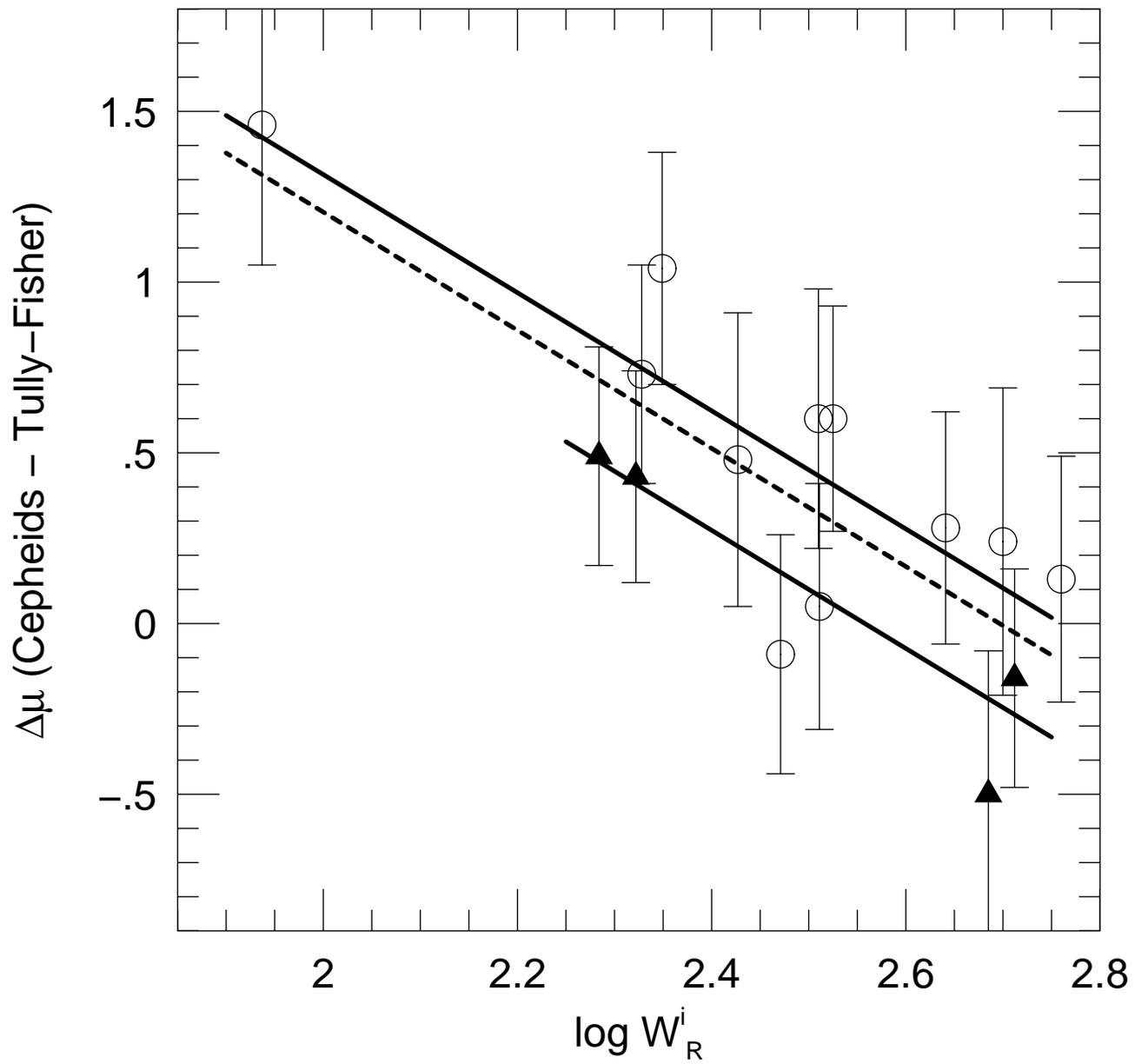
\bye